\documentclass[final,5p,times,twocolumn,authoryear]{elsarticle}
\pdfoutput=1
\usepackage{amssymb}

\usepackage{hyperref}
\hypersetup{
    colorlinks=true,
    linkcolor=myblue,
    citecolor=myblue,
    urlcolor=myblue
}
\usepackage[pagewise]{lineno}
\usepackage{newtxtext,newtxmath}
\usepackage{xcolor}
\usepackage{multirow}
\usepackage{threeparttable}
\usepackage{subcaption}
\usepackage{mwe}
\usepackage{comment}
\usepackage{adjustbox}
\usepackage{graphicx}
\usepackage{amsmath}
\usepackage{footnote}
\usepackage{tabularx}
\usepackage{graphicx}
\usepackage{soul}
\usepackage{epstopdf}

\usepackage{float}
\usepackage{lipsum}
\usepackage{pdfpages}

\usepackage[utf8]{inputenc} 
\usepackage[T1]{fontenc} 

\usepackage{amssymb}
\usepackage{lipsum}
\usepackage{newtxtext,newtxmath}
\usepackage{hyperref}

\def \swift {{\it Swift}}

\def \src{{Aql~X-1}}

\def \swiftbat{{\it Swift}/BAT}
\def \nustar{{\it NuSTAR}}
\def \maxigsc{{\it MAXI}/GSC}
\def \nicer{{\it NICER}}

\newcommand{\erg}{erg cm$^{-2}$ s$^{-1}$} 
\newcommand{\lum}{erg s$^{-1}$}

\journal{High Energy Astrophysics}

\makeatletter
\def\ps@pprintTitle{%
 \let\@oddhead\@empty
 \let\@evenhead\@empty
 \let\@oddfoot\@empty
 \let\@evenfoot\@empty
}
\makeatother

\begin{document}

\begin{frontmatter}

\title{Probing thermonuclear bursts and X-ray reflection features in \src{} during 2024 outburst}

\author[1,2]{Manoj Mandal\corref{cor1}}
\ead{manojmandal213@gmail.com}
\cortext[cor1]{Corresponding author}
\author[1]{Sabyasachi Pal}
\author[3]{G. K. Jaisawal}
\author[4]{Anne Lohfink}
\author [2]{Sachindra Naik}
\author[5]{Jaiverdhan Chauhan}

\address[1]{Midnapore City College, Kuturia, Bhadutala, West Bengal, 721129, India}
\address[2]{Astronomy and Astrophysics Division, Physical Research Laboratory, Navrangpura, Ahmedabad - 380009, Gujarat, India}
\address[3]{DTU Space, Technical University of Denmark, Elektrovej 327-328, DK-2800 Lyngby, Denmark}
\address[4]{Department of Physics, Montana State University, P.O. Box 173840, Bozeman, MT 59717-3840, USA}
\address[5]{School of Physics and Astronomy, University of Leicester, University Road, Leicester LE1 7RH, UK}

\begin{abstract}
We report the broadband timing and spectral properties of the neutron star low-mass X-ray binary \src{} during the 2024 outburst with \nicer{}, \nustar{}, and \swift{} observatories. We detected six thermonuclear X-ray bursts during the \nicer{} and \nustar{} observations, with the observed X-ray burst profiles exhibiting a strong energy dependence. The time-resolved burst spectra indicate the presence of soft excess during the burst, which can be modeled by using a variable persistent emission method ($f_a$ method) or the {\tt relxillNS} reflection model. We found that the reflection model can contribute $\sim$20\% of total emission as observed during the \nicer{} burst. The reflection and blackbody component fluxes are strongly correlated, as observed during a burst. The excess emission is possible due to the enhanced mass accretion rate to the neutron star due to the Poynting-Rodertson drag, and a fraction of burst emission may be reflected from the disk. The bursts did not show photospheric radius expansion during the peak. Moreover, we examined the burst-free accretion emission in the broadband range with \nustar{}, \nicer{}, and \swift{} at two epochs of the outburst. The persistent emission showed X-ray reflection features, which can be well modeled with the relativistic reflection model {\tt relxillCp}. The inner disk radius (R$_{in}$) is found to be nearly 22 and 10 times $\rm R_{g}$ for two observations, respectively. Assuming that the inner disk is truncated at the magnetospheric radius, the magnetic field strength at the poles of the neutron star is estimated to be $(0.6-1.9) \times 10^9$ G. 
\end{abstract}


\end{frontmatter}




\section{Introduction}
\label{intro}

 In low-mass X-ray binaries (LMXBs), matter is accreted onto the compact object through an accretion disk. The most common accretion mode in the LMXBs is the Roche-lobe overflow of material from the optical companion star \citep{Fr02}. Depending on the track in the hardness intensity diagram, the neutron star (NS) LMXBs can be classified into two categories, ‘atoll’ and ‘Z’ sources \citep{Va04}. On many occasions, the accretion disk is irradiated by the high-energy X-ray photons coming either from the hot plasma (highly energetic electrons) commonly known as corona \citep{Ha91, Ha93, Sy91}, or the boundary layer where the matter from the accretion disk makes contact with the neutron star or the surface of the NS \citep{Po01}. Currently, the precise geometry and location of the corona are not well understood. Still, it is believed to be compact and located in the inner part of the accretion disk or close to the neutron star \citep{De18}. 
 
 The unstable burning of accreted material, such as hydrogen and/or helium, on the surface of NS is responsible for a thermonuclear X-ray burst.
 During the burst, the lighter species transform into heavier elements through nuclear chain reactions \citep{Le93, St03, Sc06}. The first thermonuclear X-ray burst, known as Type-I X-ray burst, was discovered in 1976 \citep{Gr76}, which opened a new domain to study different properties of an NS. The Type-I X-ray bursts have been detected in over 100 sources to date\footnote{\url{https://personal.sron.nl/~jeanz/bursterlist.html}} \citep{Li07,Ga20}. 

The unstable fuel burning on the surface of NS results in a rapid increase in X-ray intensity \citep{Ga08}.  
Typically, the surface emission is enhanced by more than 10 times than the rest of the X-ray emission during a thermonuclear burst \citep{Ba10}. Type-I  X-ray bursts are generally of short duration and show a sharp rise (rise time of 0.5--5 s) followed by a slow exponential decay (decay time of 10--100 s) \citep{Le93}. The burst profile mostly shows single peak feature (see, e.g., \citealt{Gu22,J24}), though double-peaked profiles are observed for GX~17+2 \citep{Ku02}, 4U 1709--267 \citep{Jo04}, 4U 1636--53 \citep{Wa07}, and 4U 1608--52 \citep{Pe89, Gu21, J2019}, MXB 1730--335 \citep{Ba14}). Triple-peaked or quadruple-peaked burst profiles are much less common and were observed for 4U 1636--53 \citep{va86, Zh09, Li21}.

Thermonuclear bursts can be used to probe the emission mechanism and evolution of the temperature profile and apparent emitting area. The time-resolved spectra of the burst can be described with an absorbed blackbody model, which assumes that the NS emits like a blackbody \citep{Va78, Ku03}. Sometimes, during the peak of the Type-I X-ray burst, the luminosity can reach the Eddington limit, which leads to high radiation pressure and photosphere radius expansion (PRE) \citep{Ta84}. During PRE, the photosphere expands to a maximum value, followed by a drop in blackbody temperature at a constant luminosity level. During the rising phase, the blackbody radius increases and reaches a maximum value; after that, it decreases during the decay phase and attains the original value. At the end of the process, the photosphere becomes closer to the surface of the NS. At touchdown time, the blackbody temperature becomes maximum, and the blackbody radius drops to very low \citep{Le93, Ku03}. During the cooling phase after touchdown, the radius of a neutron star can be measured \citep{Le93, Va78, Ga08}. However, several factors, such as systematic uncertainties on the distance of the source and composition of the NS atmosphere, also contribute to the radius measurement. 

Aquila~X-1 (\src{}) is a NS LMXB discovered in 1965 \citep{Fr67}, showing a coherent X-ray pulsation \citep{Casella08} as well as burst oscillations at 550 Hz \citep{Zh98, Ga08, Bi19}. It shows frequent outbursts about once a year \citep{Ca14, Gu14}. \src{} has a K-type donor at a distance of nearly 6 kpc \citep{Ma17}. Several efforts have been made to study thermonuclear bursts in \src{} using data from various observatories \citep{Gu22, Ke18, Ga08}. Earlier, reflection from the accretion disk was also reported in the persistent spectra of \src{} \citep{Ki16, Lu17}. 

Irrespective of the origin of thermal X-ray photons, the outer layers of the accretion disk also reprocess the irradiating hard X-ray photons and re-emit the reprocessed emission. The energy spectrum corresponding to these re-emitted photons shows various peculiar features, including an iron (Fe K$\alpha$) emission line at 6.4 keV, a dip around 10 keV representing photoelectric absorption of low energy X-ray photons, and a Compton back-scattering hump above 20 keV superimposed on a continuum identified as ``reflection'' spectrum \citep{Ge91,Fa89, Fa00, Mi07}. The Fe K$\alpha$ emission line is broadened due to the Doppler effect, general relativistic effects, and the scattering of the X-ray photons with the hot inner flow \citep{Fa89, Fa00, Mi07}. These reflection features have been reported in both black holes (e.g., \citep{Mi02a, Mi02b, Bh19a, Bh19b}) and NS (e.g., \citep{Bh07, Ca08}) LMXBs, which suggests that even though the masses of the accretors are different, both represent similar accretion geometries where line-emitting region reaches close to the compact object.

In 2024, an X-ray outburst was detected from \src{} using the {\it Einstein} probe \citep{Liu24}. Following X-ray activity, the source was followed up in multi-wavelength ranges \citep{Ma24a, Ma24b, Ru24, Ro24, Gr24, Le24, Mi24}. In this paper, we report the results of the spectral and temporal study of \src{} during the 2024 outburst using data from \nustar{}, \swift{}, and \nicer{} observatories. We detected 6 thermonuclear bursts from \src{} with \nicer{} and \nustar{}. Temporal and spectral studies are performed to understand the broadband characteristics of each burst using combined \nicer{} and \nustar{} observations. The broadband spectral analysis showed evidence of an X-ray reflection feature, which is used to probe the accretion disk properties. The X-ray reflection spectral study is performed at two different phases of the outburst using \nustar{}, \swift{}, and \nicer{} observations to understand the evolution of spectral parameters and disk properties. This paper is organized as follows. The observations and data reduction procedure are described in Section \ref{obs}. Section \ref{res} summarizes the results of the spectral and timing analysis, followed by a discussion of the results in Section \ref{dis}. The summary and conclusions of the work are presented in Section \ref{con}.

 \begin{figure}
\centering{
\includegraphics[width=\columnwidth]{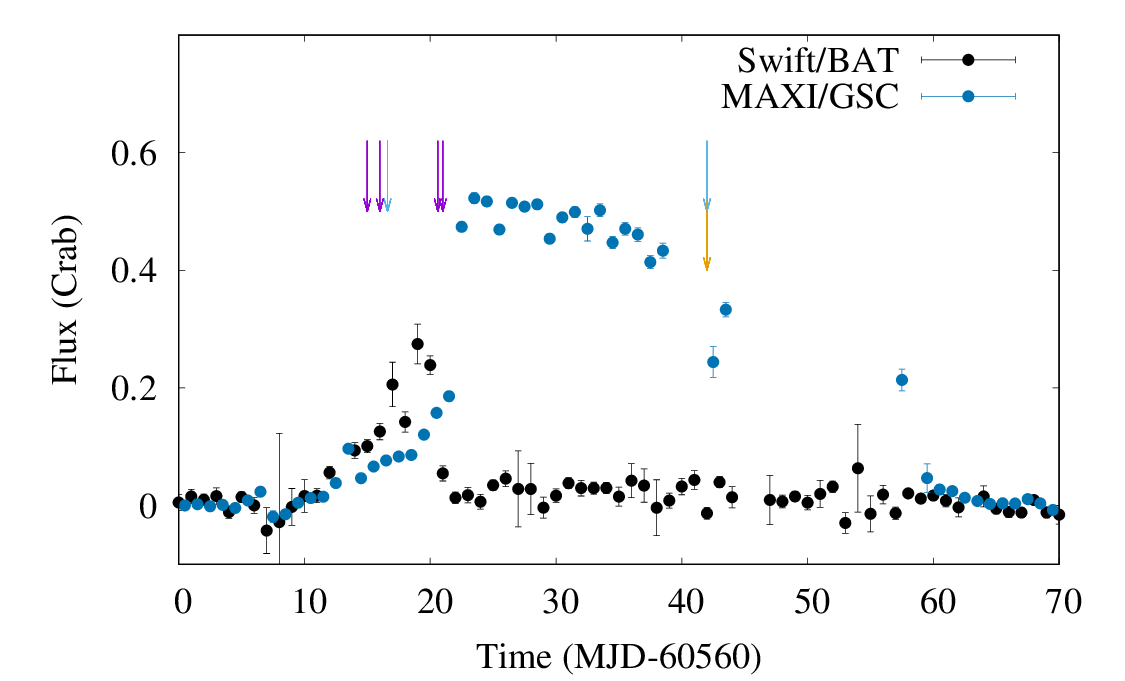}
\caption{The \swiftbat{} (15--50 keV) and \maxigsc{} (2--20 keV) light curves of \src\ during its 2024 outburst. The blue, purple, and orange arrows represent the time of \nustar{}, \nicer{}, and \swift{} observations, respectively.}
\label{fig:BAT}}
\end{figure}

\begin{table*}
\centering
\caption{Summary of observations and details of X-ray bursts detected from \src{} using {\it NuSTAR} and {\it NICER} observations. The reported unabsorbed peak flux is estimated in the 0.1--100 keV range (in units of 10$^{-8}$ erg cm$^{-2}$ s$^{-1}$).}
\begin{tabular}{ccccccccc} 
\hline
	
	 & Obs. T$_\textrm{start}$ & Burst T$_\textrm{start}$      & No. of  & Obs. ID & Exposure  & $\frac{peak}{persistent}$ & Peak count & Peak flux\\
		   &   (MJD)      &  (MJD)  & bursts   &       & (ks) &       &  (c s$^{-1}$) &          \\
\hline

{\it NuSTAR} &  60576.59 & 60576.68687& 3 & 91001338002 (obs-1) & 19 & 14.3      & 1150$\pm$72 & $4.5\pm0.2$ \\
 &   &  60576.80556 &  &  & &  14.8     & 1180$\pm$77  & $4.7\pm0.2$ \\
 &   &  60577.01840 &  & &  & 14.8      & 1187$\pm$79  & $4.7\pm0.2$  \\
& 60602.11 & --  &   -- & 91001345002 (obs-2) & 20 &      -- &  --  & --  \\
\hline

{\it NICER}& 60575.01 & 60575.85534  &   1 & 7050340103 (obs-1) & 5.9 & 7.5       & 2812$\pm$53  & $6.6\pm0.6$ \\
& 60576.04   &   -- &-- & 7050340104 (obs-2) & 4.1 & --      & --  & --  \\
& 60580.68   & 60580.89451&  1 & 7050340108 (obs-3) & 3.2 & 3.5     & 2313$\pm$48  & $4.0\pm0.2$ \\
& 60581.07   &  60581.08520& 1 & 7675010106 (obs-4) & 1.9 & 3.5      & 3252 $\pm$57  & $4.5\pm0.6$  \\
\hline
{\it Swift}& 60602.27& --   &   -- & 00089981001 & 2.0 & --       & --  & -- \\
\hline
\label{tab:log_table_burst}
	\end{tabular}
\end{table*}

\section{Data analysis and methodology}
\label{obs}
We used data from \nicer{}, \swift{}, and \nustar{} during the 2024 outburst of \src{}. The data are reduced using {\tt HEASOFT} version 6.31.1. We utilized final data products (light curves) provided by {\it MAXI} \citep{Ma09} and the Burst Alert Telescope ({\it BAT}) on board the Neil Gehrels Swift Observatory \citep{Kr13} missions.

\subsection{{\it NuSTAR} observation}

The Nuclear Spectroscopic Telescope Array (\nustar{}) is an imaging X-ray telescope consisting of two identical detectors that are co-aligned and working in 3–79 keV energy band \citep{Ha13}. \nustar{} observed \src{} twice under the Director’s Discretionary Time (DDT) proposal on MJD 60576.6 and MJD 60602.1 for effective exposures of around 20 ks each. The details of the observation are presented in Table \ref{tab:log_table_burst}. For data reduction, we used the standard NuSTAR data analysis software (NUSTARDAS v2.1.2) distributed with HEASOFT v 6.31.1 with the latest {\tt CALDB} version 20240325. We used the {\tt nupipeline} task to filter the event files. We used a circular region of radius 120 arcsec centered on the source location to extract the source events. A similar circular region was defined away from the centre of the source region to create background events. Finally, we used the {\tt nuproducts} task to generate the energy spectrum, the auxiliary response file, and the response matrix file for both detectors. The light curves are background corrected using the {\tt lcmath} task. Further, we simultaneously modeled the energy spectra from both detectors to minimize the systematic effects. 

\subsection{{\it NICER} observation}

The Neutron Star Interior Composition Explorer (\nicer{}) is a soft X-ray non-imaging telescope mounted on the International Space Station, which operates in the energy range of 0.2--12 keV \citep{Ge16}. \nicer{} monitored the 2024 outburst of \src{} with high cadence starting from 2024 September. For this work, we used only 3 \nicer{} observations of the source during which thermonuclear bursts were detected. The details of the observation are presented in Table \ref{tab:log_table_burst}. In addition, another \nicer{} observation (Obs-2) is used to probe the broadband spectral properties, which was close to the \nustar{} observation (obs-1). Initially, we processed the raw data using the tool {\tt NICERDAS} in {\tt HEASOFT}. We generated clean event files from the raw data after applying standard filtering and calibration tool {\tt nicerl2}. The barycentric correction was performed using the {\tt barycorr} tool. Finally, we used the {\tt XSELECT} package to create light curves and spectra from the clean event files. We used {\tt CALDB} version xti20221001 for analyzing the \nicer{} data. The background spectrum was generated using the {\tt nibackgen3C50} tool \citep{Re22}. The response and ancillary response files for spectral analysis were created using {\tt nicerrmf} and {\tt nicerarf} tools.

\subsection{{\it Swift} observation}

The Swift/X-ray Telescope (XRT; \citet{Ge04}) observed \src{} on 19 October 2024, which was close to the \nustar{} obs-2. The observation details are summarized in Table \ref{tab:log_table_burst}. We used this \swift{} observation to probe the broadband spectral study of the source along with \nustar{}. The source was observed in the Windowed Timing (WT) mode. Initially, we used standard {\tt xrtpipeline} version 0.13.7 for filtering, screening, and reducing the Swift/XRT data. To avoid the photon pile-up effect, an annulus region of radii 10 arcsec and 50 arcsec is selected for spectral study. The background spectrum is generated using a circular region of 60 arcsec. 
We used {\tt FTOOL XSELECT }(V2.5b) to extract the source and background spectra. The exposure map and ancillary files are created using the {\tt xrtexpomap} and {\tt xrtmkarf}, respectively. For further spectral analysis, the resulting files and proper RMF files from the recently updated CALDB are used. The spectrum is grouped for a minimum of 25 counts/bin using {\tt grppha} task.

\begin{figure*}
\centering{
\includegraphics[width=8.5cm]{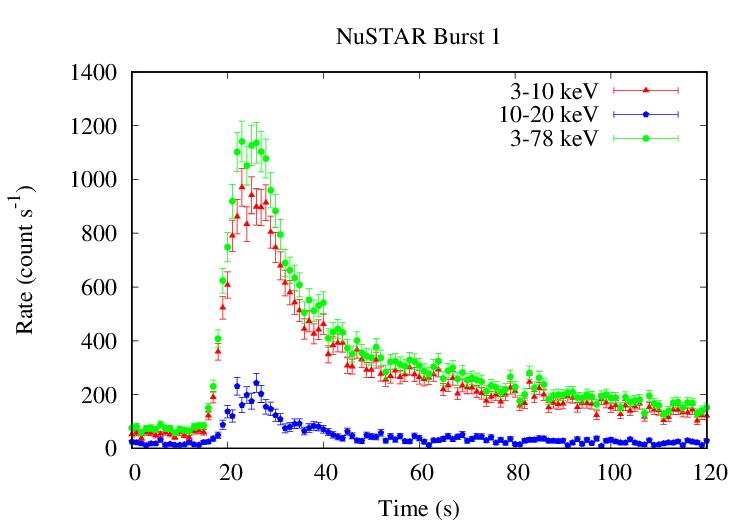}
\includegraphics[width=8.5cm]{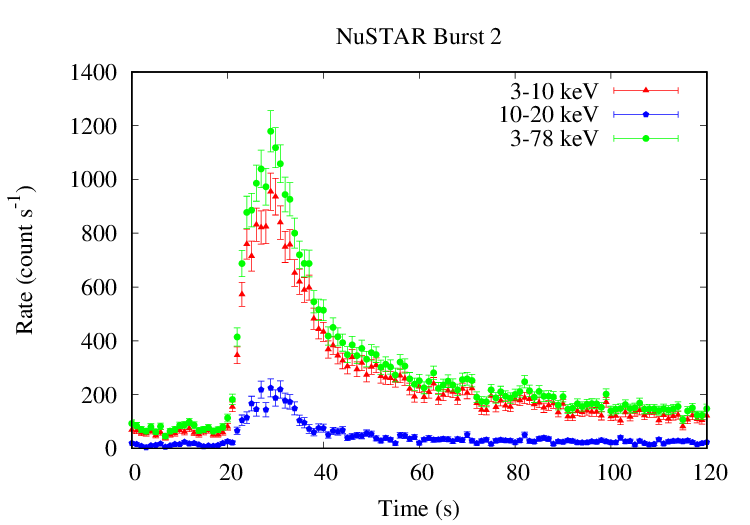}
\includegraphics[width=8.5 cm]{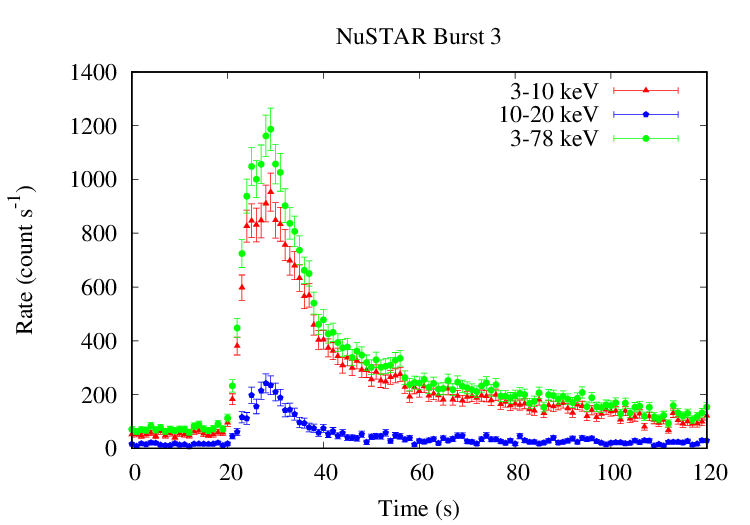}
\includegraphics[width=8.5 cm]{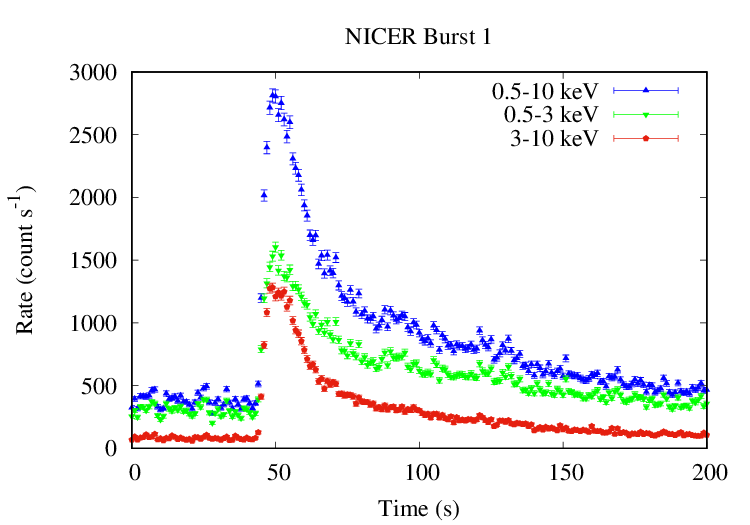}
\includegraphics[width=8.5 cm]{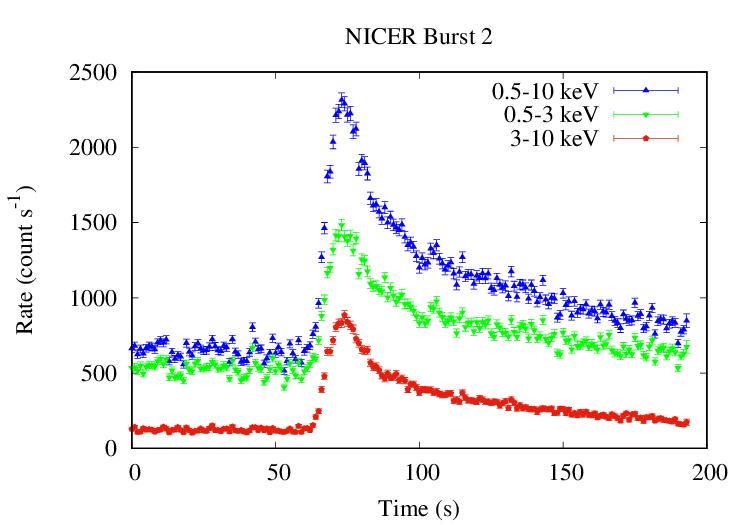}
\includegraphics[width=8.5 cm]{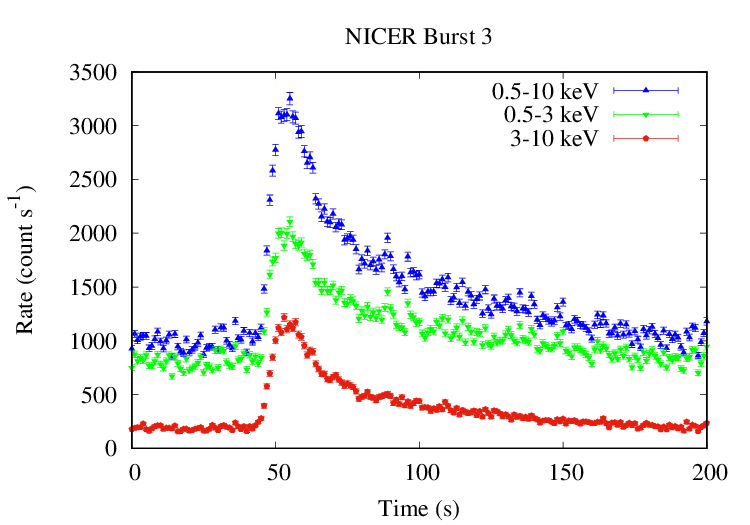}
\caption{Energy-resolved burst profiles are shown for {\it NICER} and {\it NuSTAR}/FPMA (obs-1) observations. The top left and right-hand side panels show energy-resolved burst profiles for \src{} using {\it NICER} data with a 1-s binned light curve, and the rest of the bursts are observed using {\it NuSTAR}. Three thermonuclear bursts are detected from \src{} in the {\it NuSTAR} 1~s binned light curves. The bursts are detected significantly up to an energy of 20 keV using {\it NuSTAR}.}
	 \label{fig:burst_profile}}
\end{figure*}

\section{Results}
\label{res}
 \src{} went through an outburst in 2024 as observed by {\it MAXI} and \swiftbat{}. Several follow-up observations have been conducted to study the multi-wavelength behavior of the source \citep{Ma24a,Ma24b,Ru24,Ro24, Gr24, Le24, Mi24}. We utilized the \textit{Swift}/BAT telescope's high-energy X-ray coverage (15-50\,keV) and daily monitoring data to probe the spectral state of the source and the evolution of the outburst. Fig.~\ref{fig:BAT} shows the outburst from \src{} during 2024 September--October as seen with {\it MAXI} and {\it BAT}. The maximum flux reached by the source was $\sim$0.3 crab as observed with {\it Swift}/BAT in the 15--50 keV range. The epochs of the \nustar{}, \swift{}, and \nicer{} observations are indicated with arrows (see Fig.~\ref{fig:BAT}).


\begin{table*}
\centering
 \caption{Best-fitting spectral parameters [{\tt XSPEC} model $\texttt{TBabs}\times(\texttt{bbodyrad}\,\times\,\texttt{po})$] of the pre-burst {\it NICER} (0.5--10 keV) and {\it NuSTAR} (3--30 keV) spectra of \src{}.}
\begin{tabular}{llcccccc}
\hline
Components & Parameters & \nicer{} & \nicer{} & \nicer{} & \nustar{} & \nustar{} & \nustar{} \\
 &  & TNB1 & TNB2 & TNB3 & TNB1 & TNB2 & TNB3 \\
\hline
{\tt tbabs} & N$_{H}$ & 0.50$\pm0.02$ & 0.50$\pm0.02$ & 0.48$\pm0.01$ & 0.5$^{f}$ & 0.5$^{f}$ & 0.5$^{f}$ \\
\hline
{\tt bbodyrad} & kT$_{bb}$(keV) & 1.0$\pm$0.1 & 1.8$\pm$0.1 & 2.0$\pm$0.1 & 1.3$\pm$0.1 & 1.4$\pm$0.1 & 1.3$\pm$0.4 \\
  & Norm. & 27.5$\pm$10 & 6.5$\pm$2.5 & 7.9$\pm$2.5 & 14.2$\pm$5.0 & 11.3$\pm$5.0 & 7.6$\pm$8.5\\
\hline
{\tt power-law} & $\Gamma$ & 1.47$\pm0.05$ & 1.73$\pm0.08$ & 1.73$\pm0.06$ & 1.62$\pm0.06$ & 1.70$\pm0.06$ & 1.72$\pm0.09$ \\
  & Norm. & 0.22$\pm0.01$ & 0.46$\pm0.01$ & 0.70$\pm0.02$ & 0.31$\pm0.06$ & 0.38$\pm0.06$ & 0.44$\pm0.11$ \\
\hline
 & F$_{Total}$  & 2.81$\pm 0.03$ & 4.18$\pm 0.03$ & 6.55$\pm 0.03$ & 5.78$\pm 0.12$ & 5.60$\pm 0.08$ & 5.88$\pm 0.13$ \\
 & Luminosity  & 1.20$\pm$ 0.01 & 1.80$\pm$ 0.01 & 2.84$\pm 0.01$ & 2.50$\pm 0.05$ & 2.41$\pm 0.03$ & 2.53$\pm 0.05$ \\
\hline
 & $\chi^{2}$/dof & 506/526 & 670/618 & 685/683 & 587/580 & 635/586 & 329/378 \\
\hline
\label{tab:tab_preburst}
\end{tabular}\\
{\bf $f$} indicates the frozen parameters. 
{\bf }: The reported errors are of 90\% significant. All fluxes are unabsorbed (1--79 keV) in unit of 10$^{-9}$ ergs cm$^{-2}$ s$^{-1}$, the X-ray luminosity in the units of 10$^{37}$ erg s$^{-1}$, $N_H$ is in the unit of $\rm 10^{22}~ cm^{-2}$.
\end{table*}

\subsection{Burst light curves and burst profiles}
We detected a total of 6 thermonuclear X-ray bursts using \nicer{} and \nustar{} observations during the 2024 outburst of \src{}.
In the first {\it NuSTAR} observation (obs-1), three X-ray bursts are detected, and during the second \nustar{} observation (obs-2), no burst was observed. The average peak count rate during these \nustar{} bursts was around 1170 counts s$^{-1}$, and the peak-to-persistent ratio was around 15. The average persistent count rate was $\sim$45 counts s$^{-1}$ for the three \nustar{} bursts. During \nicer{} monitoring of \src{}, three thermonuclear bursts were also observed on different days of the outburst. The summary of detected bursts is given in Table~\ref{tab:log_table_burst}. The peak count rates for the detected X-ray bursts with \nicer{} were around 2812 counts s$^{-1}$, 2313 counts s$^{-1}$, and 3252 counts s$^{-1}$, respectively. The peak-to-persistent ratio for burst-1 was nearly 7.5, whereas for burst-2 and burst-3, the peak-to-persistent ratio was around 3.5. The average persistent count rate was nearly 350 counts s$^{-1}$ (TNB1), 600 counts s$^{-1}$ (TNB2), and 1000 counts s$^{-1}$ (TNB3) for \nicer{} bursts.

The burst profiles of \src{} are investigated using \nustar{} and \nicer{} data. The burst profiles are generated in 0.5--3 keV, 3--10 keV, and 0.5--10 keV ranges for \nicer{}, and in 3--10 keV, 10--20 keV, and 3--78 keV ranges for \nustar{}. The energy-resolved burst profiles from the \nustar{} and \nicer{} observations are shown in Fig.~\ref{fig:burst_profile}. The burst profiles showed significant evolution with energy. A longer tail is visible in the burst profiles at lower energies. The X-ray bursts are detected up to 20 keV with \nustar{}. The peak count rate of the \nustar{} burst profile is reduced by a factor of $\sim$5 in the higher energy band (10--20 keV range) compared to the lower energy band (3--10 keV range). 


\begin{figure*}[t]
    \includegraphics[width=0.68\columnwidth]{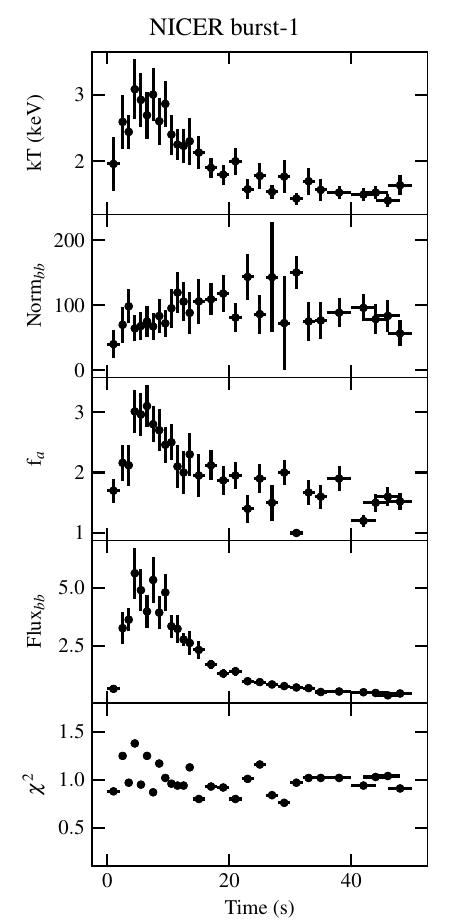} 
  \includegraphics[width=0.68\columnwidth]{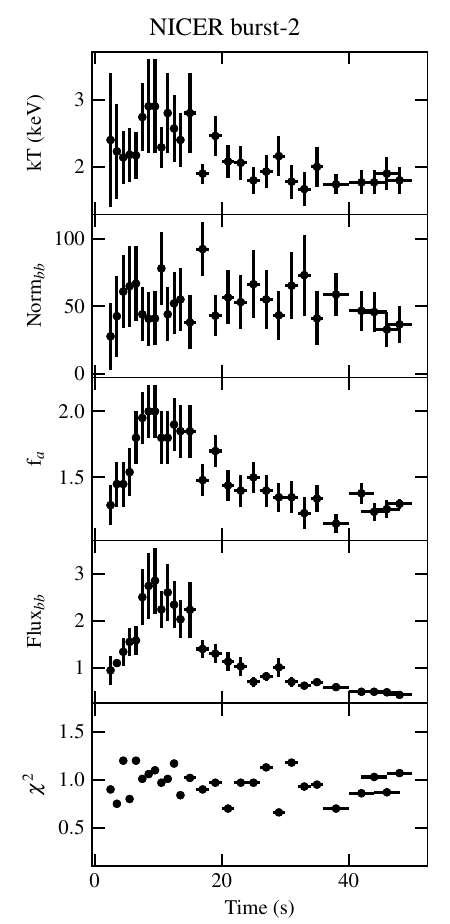}
    \includegraphics[width=0.68\columnwidth]{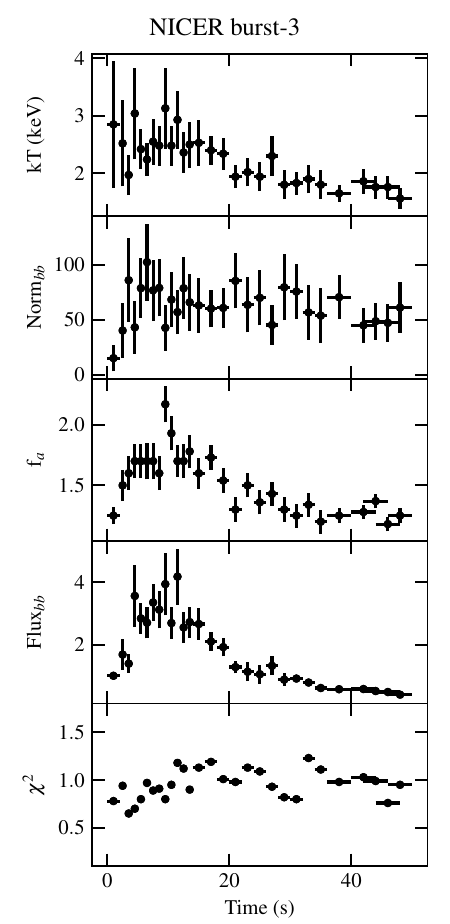}

\caption{Evolution of various spectral parameters of \src\ during the evolution of the thermonuclear burst with \nicer: blackbody temperature (kT; Top panel), blackbody normalization (second panel), $f_a$ parameter (third panel), blackbody flux (0.1--100 keV range) in the unit of 10$^{-8}$ erg cm$^{-2}$ s$^{-1}$ (fourth panel), and $\chi^{2}$/dof (bottom panel). The X-axis presents the time from the rise of the burst.}
    \label{fig:fig_time_Resolved_nicer}
\end{figure*}

\begin{figure*}[t]
    \includegraphics[width=0.68\columnwidth]{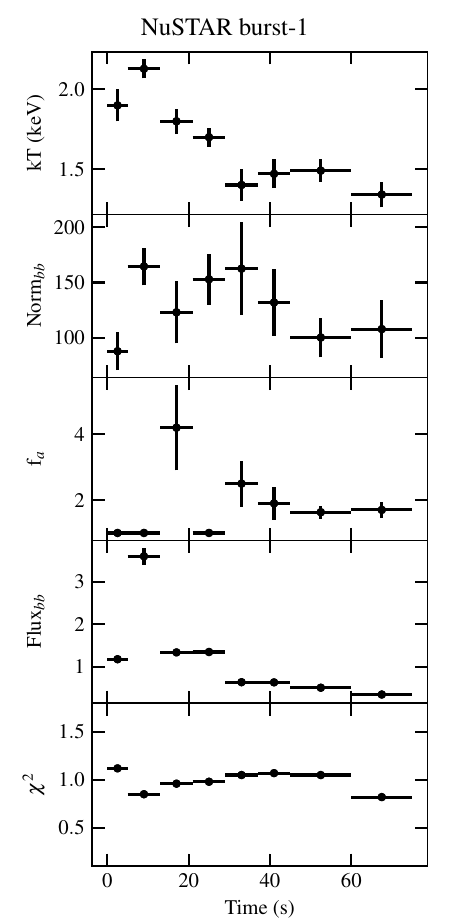} 
   \includegraphics[width=0.68\columnwidth]{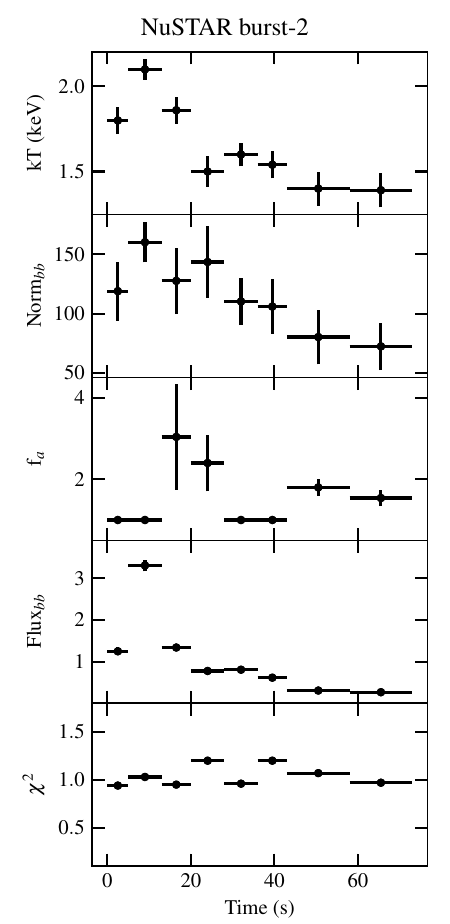}
      \includegraphics[width=0.68\columnwidth]{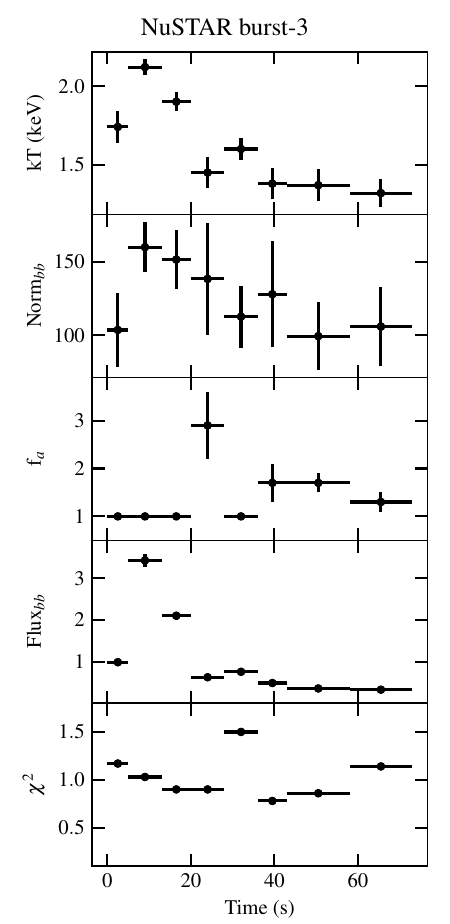}

\caption{Evolution of various spectral parameters of \src\ during the evolution of the thermonuclear burst with \nustar{} (obs-1): blackbody temperature (kT; Top panel), blackbody normalization (second panel), $f_a$ parameter (third panel), blackbody flux (0.1--100 keV) in a unit of 10$^{-8}$ erg cm$^{-2}$ s$^{-1}$ (fourth panel), and $\chi^{2}$/dof (bottom panel). The X-axis presents the time from the rise of the burst.}
    \label{fig:fig_time_Resolved_nustar}
\end{figure*}

\subsection{Time-resolved burst spectroscopy}
\label{time_resolved_spectroscopy}
To probe the dynamic evolution of the X-ray bursts, time-resolved spectroscopy of each burst is performed to investigate the variation of different spectral parameters during the bursts. The pre-burst spectrum of each burst, either from \nicer{} or \nustar{} observation, is modeled in {\tt XSPEC} \citep{Ar96}. The pre-burst \nustar{} spectra are obtained for a time bin of $\sim$300 s. The pre-burst spectra are modeled with a power law continuum and a blackbody ({\tt bbodyrad}) model. The {\tt TBABS} \citep{Wilms2000} model is used to account for the interstellar medium absorption. The best-fit pre-burst spectral fitting parameters are summarized in Table~\ref{tab:tab_preburst}. The hydrogen column density ($N_H$) from the pre-burst spectra was $\sim 0.5 \times 10^{22}$ cm$^{-2}$. 

To perform the time-resolved spectral study for each burst with \nicer{} (0.5--10\,keV), a total of 30 spectra (P1 to P30) are generated for each burst. The time bin of 1 s is used to create the \nicer{} spectra up to the P15 segment, and then the time bin is increased gradually to 2 s (for the segments P16--P25) and 4 s (for the segments P26--P30). To optimize the signal-to-noise ratio during the burst decay part, the time bin is extended from 2\,s to 4\,s. For the time-resolved spectral analysis with \nustar{}, each burst is divided into 8 segments (P1--P8). The burst rising part (P1) is for 5\,s segment, the broad flat peak (P2) is for 8\,s, and P3--P6 is for 8 s each. The bin time is increased to 15\,s during the decay part.
  
The time-resolved spectra for all six bursts are described with an absorbed blackbody model ({\tt bbodyrad}), and the $f_a$ method is used to scale the pre-burst emission \citep{Worpel2013}. In the $f_a$ method, the persistent emission is allowed to vary during the burst, and a scaling factor is used to accommodate any excess observed during the X-ray burst. The mass accretion rate is supposed to increase during the burst, and the $f_a$ method provides a significant improvement ($P$<0.05 in F-test) modelling the results when compared to the constant persistent emission approach. Consequently, the $f_a$ method provides an acceptable fit ($\chi^2$/dof<1.3) for most of the spectra. We performed the F-test to check the significance of the $f_a$ component. If the chance probability of the decrease in $\chi^2$ values is lower than 5\%, then the $f_a$ modeling approach is used; otherwise, the $f_a$ value is fixed at 1. For the peak of \nicer{} bursts, the F-test results provide $F$-statistic value of $\sim$55, 61, and 212 for three bursts, respectively, with the F-test probability less than $10^{-11}$. Similarly, it is checked for all spectra, and in all \nicer{} time-resolved burst spectra (except 1 segment in \nicer {} burst-1), the $f_a$ model was statistically significant. The hydrogen column density is fixed to the best-fit pre-burst value $\sim 0.5\times 10^{22}$ cm$^{-2}$ during \nicer{} and \nustar{} time-resolved spectral fitting. 

The evolution of spectral parameters during the bursts for \nicer{} observations is shown in Fig.~\ref{fig:fig_time_Resolved_nicer}. The blackbody flux (in 0.1--100 keV range) of $\sim 5 \times 10^{-8}$ erg cm$^{-2}$ s$^{-1}$, $\sim 3 \times 10^{-8}$ erg cm$^{-2}$ s$^{-1}$, and $\sim 4 \times 10^{-8}$ erg cm$^{-2}$ s$^{-1}$, are observed during the peak of \nicer{} burst 1, burst 2, and burst 3, respectively. No sign of PRE is detected during any of the bursts. The $f_a$ values of maximum 3, 2, and 2 are reached for burst-1, burst-2, and burst-3, respectively. During the \nustar{} time-resolved spectral analysis, we used the same model as used for \nicer{} bursts. The evolution of different spectral parameters during the bursts observed with \nustar{} is presented in Fig.~\ref{fig:fig_time_Resolved_nustar}. The blackbody temperature and normalization reached 2.2 keV and 150, respectively, during the peak of the burst. The blackbody flux (0.1--100 keV range) reached a maximum value of $\sim3.5 \times 10^{-8}$ erg cm$^{-2}$ s$^{-1}$, $\sim2.9 \times 10^{-8}$ erg cm$^{-2}$ s$^{-1}$, and $\sim3.1 \times 10^{-8}$ erg cm$^{-2}$ s$^{-1}$ for burst-1, burst-2 and burst-3, respectively. The $f_a$ parameter varied between 1.5 and 4 during the \nustar{} bursts. The peak flux during the burst is mentioned in Table~\ref{tab:log_table_burst}. During the peak of the bursts, the source luminosity reached 0.5 and 0.7 times the Eddington limit for the \nustar{} and \nicer{} bursts, respectively, ( assuming the Eddington limit of $L_{Edd} = 3.8 \times 10^{38}$ erg s$^{-1}$ for a NS of mass 1.4 $M_\odot$ and radius 10 km at a distance of $\sim$6 kpc \citep{Ku03}).

\begin{figure}[t]
    \includegraphics[width=\columnwidth]{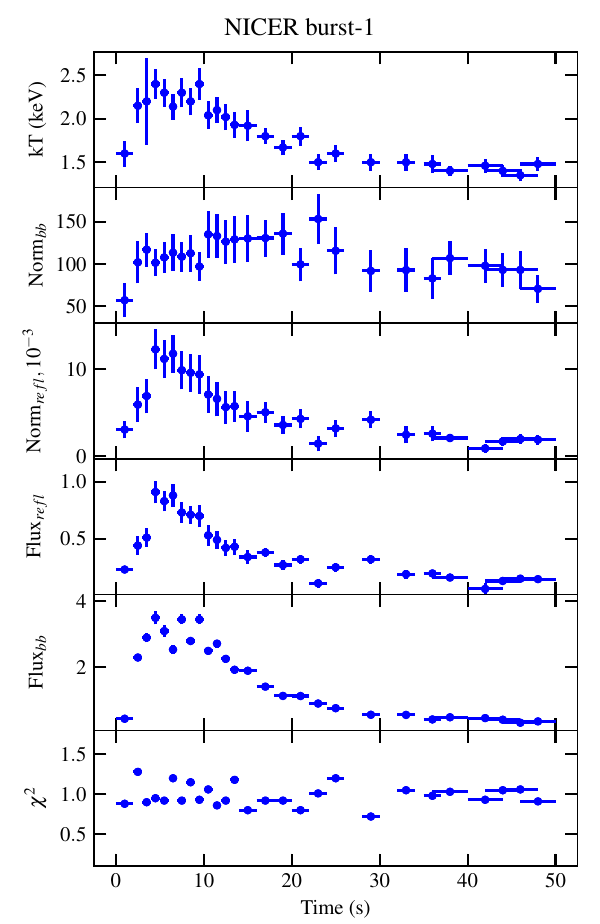} 
\caption{Evolution of various spectral parameters of \src\ during the thermonuclear burst with \nicer: blackbody temperature (kT; Top panel), blackbody normalization (second panel), normalization of {\tt relxillNS} (third panel), reflection component flux  (fourth panel), blackbody component flux (fifth panel) and $\chi^{2}$/dof (bottom panel). Measured fluxes are in the energy range of 0.1--100 keV in the unit of 10$^{-8}$ erg cm$^{-2}$ s$^{-1}$. The X-axis presents the time from the rise of the burst.}
    \label{fig:nicer_reflection}
\end{figure}

\begin{figure}
\centering{
\includegraphics[width=\columnwidth]{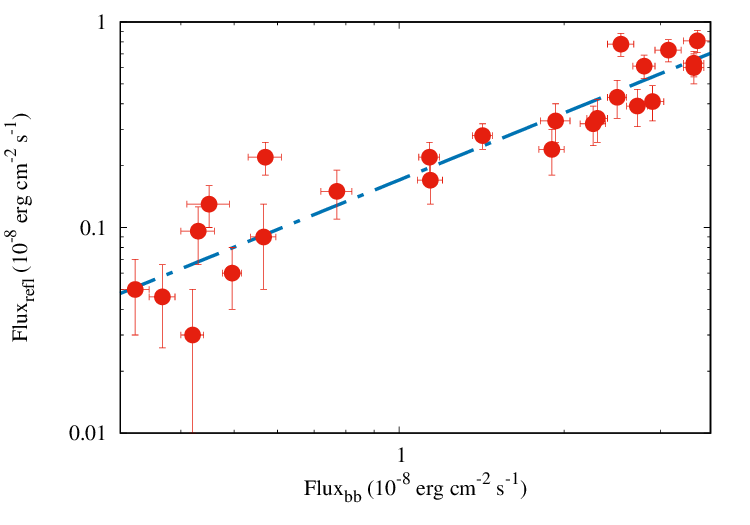}
\caption{Correlation between the blackbody and reflection component flux during the \nicer{} burst-1. 
}
\label{fig:correlation}}
\end{figure}
 
\subsubsection{Disk reflection during burst: burst-disk interaction}
The deviation of the burst spectrum from a pure blackbody model can be probed with the disk reflection model. We used the latest relativistic reflection model {\tt relxillNS} to investigate the reflection feature in the burst spectra of Aql~X-1. In this model, the photoionized disk is illuminated by the blackbody spectrum from the neutron star \citep{Garcia2022}. The \nicer{} burst spectra can be modeled with {\tt tbabs $\times$(bbodyrad1+relxillNS+powerlaw+bbodyrad2)}, where {\tt bbodyrad1}, {\tt relxillNS}, and ({\tt powerlaw+bbodyrad2}) represent the component of the burst, the reflection of burst photons from the accretion disk and persistent emission, respectively. During the fitting of burst spectra with the above reflection model, the preburst model parameters are fixed at the preburst values obtained during the earlier model fitting (Section~3.2). The {\tt relxillNS} model parameters are: $q1$ and $q2$ are the inner and outer indices, $i$ is the disk inclination, $R_{in}$ and $R_{out}$ are the inner and outer disk radii, $a$ is the dimensionless spin parameter, $A_{Fe}$ is the iron abundance, $kT_{bb}$ represents the input blackbody temperature, log $\xi$ is the disk ionization parameter and $logN$ is the disk density. The emissivity profile of the accretion disk is defined as a broken power law with three governing parameters: index1 (q1), index2 (q2), and break radius ($\rm R_{br}$). Assuming a single emissivity profile ($r^{-q}$), The emissivity index is fixed at q1 = q2 = 3, which eliminates the significance of $R_{br}$. As the reflection model contains many parameters, it is challenging to constrain all parameters simultaneously in short intervals. Therefore, during the fitting, the temperature of the input blackbody spectrum is tied with the {\tt bbodyrad1} component, i=30$^{\circ}$, logN=17 cm$^{-3}$, $A_{Fe}$=5, the inner and outer disk radii are fixed at $\rm R_{ISCO}$, and 400 $R_g$, respectively. Initially, the disk ionization parameter (log $\xi$) was allowed to vary. We found that the variation in the ionization parameter is not significant during time-resolved spectroscopy. Therefore, we have fixed this parameter to a value of log $\xi$=3.2 erg cm s$^{-1}$. The reflection fraction is fixed at -1, and the normalization of the {\tt relxillNS} is allowed to vary during the burst. The F-test also indicates that the {\tt relxillNS} model is statistically significant ($F$ value and probability of 20.5 and  $2.5\times10^{-10}$, respectively, during the peak). Therefore, we fitted the burst spectra with the reflection model for the \nicer{} observation, for which the $f_a$ model was used. As the burst parameters showed a similar type of evolution during the \nicer{} bursts, we showed the evolution of burst parameters with {\tt relxillNS} model for the \nicer{} burst-1 in Fig.~\ref{fig:nicer_reflection}. It can be seen that the contribution of the reflection component increases with the increase in the burst flux. During the burst peak, the flux (0.1--100 keV range) of the {\tt relxillNS} and blackbody components reached a maximum value of $\sim10^{-8}$ erg cm$^{-2}$ s$^{-1}$ and $\sim$ 3.5 $\times 10^{-8}$ erg cm$^{-2}$ s$^{-1}$, contributing about 20\% and 74\% towards the total flux, respectively. The blackbody temperature and flux are comparatively lower than the values obtained from the $f_a$ method.

In addition, we investigated the correlation between the flux contributions from the reflection component (F$_{refl}$) and blackbody component (F$_{bb}$) during the burst. In doing so, the contribution from the persistent reflection component is subtracted from the burst reflection flux. In Fig.~\ref{fig:correlation}, we plotted the flux of the reflection component with the blackbody flux obtained from the time-resolved spectroscopy of the \nicer{} burst-1. The flux of the reflection component is strongly correlated with the flux of the blackbody component, as shown in the figure. We fitted it with a function $F_{refl} = k~F_{bb}^{\alpha}$. The best-fit result provides $k$ = 0.17$\pm$0.01 and $\alpha$ = 1.08$\pm0.09$.

\subsection{Analysis of persistent spectra: Reflection}
\label{spectral}

In addition to X-ray bursts, we studied the broadband persistent emission at two epochs of the outburst, using data from \nustar{}, \nicer{}, and \swift{} observations. We attempted to probe the continuum emission and reflection from the system. We excluded the data during the thermonuclear bursts for \nustar{} obs-1 and \nicer{} obs-1 to perform the reflection spectral study for epoch 1. For the second \nustar{} observation (obs-2), we used data up to 30 keV as the data beyond this are background-dominated. For the broadband spectral fitting, data from \swift{}/XRT are also used. To account for the uncertainties related to the cross-instrument calibration, we include an extra constant multiplier factor using the constant model in {\tt XSPEC}. This constant was set to a value of 1 for \nicer{} and \swift{}/XRT, while for FPMA/FPMB, it was allowed to vary. To model the absorption along the line of sight due to the Galactic neutral hydrogen, we used {\tt TBabs} model \citep{Wilms2000} provided in the {\tt XSPEC}. We used the most recent abundance model {\tt wilm} incorporated in the {\tt XSPEC} for the initial input abundance in the {\tt TBabs} model.


\renewcommand{\arraystretch}{1.2}
\begin{table*}
	\centering
	\caption{Broadband spectral fitting parameters of \src{} for different model combinations. The best-fit is obtained with spectral model \textcolor{black}{M4: \texttt{cons$\times$tbabs$\times$(nthcomp+diskbb+relxiilCp)}}. All errors are calculated using the MCMC in {\tt XSPEC} and are 90\% significant.  
 }
	\label{tab:fitstat2}
	\resizebox{\linewidth}{!}{
	\begin{tabular}{lccccccccc} 
\hline
				
 Model 	&	 Parameters	&	\multicolumn{4}{c}{ Obs-1}& Obs-2	\\							&		&	 Model M1	&	 Model M2	&	 Model M3 	&	 Model M4  	& Model M4	&  & &\\
\hline																			
    \texttt{tbabs} 	&	 N$_H$ ($10^{22}$ cm$^{-2}$) 	&	 $0.30 \pm 0.01$ 	&	 $0.30\pm0.01$ 	&	 $0.4\pm0.01$ 	&	 $0.39\pm 0.03$ & $1.02_{-0.05}^{+0.03}$ & \\										
    \texttt{nthcomp} 	&	 kT$_{\rm e}$ (keV) 	&	$ 29.1 \pm 2.5 $	&	 $14.8 \pm 0.5$ 	&	 $14.8 \pm 0.5$ 	&	 $17.3_{-0.6}^{+1.1}$ & $2.6\pm1.0$& \\
     &	 kT$_{\rm seed}$ (keV) 	&	$ 1.13 \pm 0.01 $	&	 $1.63$ 	&	 $1.49$ 	&	 $1.54$ & $1.23$ & \\
	&	Photon index ($\Gamma$)	&	$1.88 \pm	0.01$ &	 $1.75 \pm 0.01$ 	&	 $1.75 \pm 0.01$ 	&	 $1.80_{-0.005}^{+0.01}$ 	& $2.412\pm0.0001$ & \\


   \texttt{diskbb} 	&	 $kT_{\rm in}$ (keV) 	&	$-$ &	 $1.63 \pm 0.02$ 	&	 $1.49 \pm 0.02$ 	&	 $1.54\pm0.04$ & $1.23_{-0.04}^{+0.1}$ & \\
    &	Norm	&	 $-$ 	&	 $4.9 \pm 0.2$ 	&	 $4.9 \pm 0.2$ 	&	 $4.2 \pm 0.2$ &	$18.4_{-8.1}^{+12.1}$	&\\
  
\texttt{Gaussian} 	&	 $E_{\rm line}$ (keV) 	&	$-$	&	 $-$ 	&	 $6.44 \pm 0.06$ 	&	 $-$ &  $-$ & \\

&	 $\sigma$ (keV)	&	$-$ &	 $-$ 	&	 $0.62 \pm 0.08$ 	&	 $-$ 	& $-$ & \\

&	 norm (10$^{-3}) $	&	$-$	&	 $-$ 	&	  $1.77 \pm 0.25$ 	&	 $-$ 	&	$-$ & \\
 
 \texttt{relxillCp} 	&	 Incl (deg) 	&	$-$	&	 $-$ 	&	 $-$ 	&	 $33.1_{-3.2}^{+5.8}$ 	&	$19.5_{-2.7}^{+3.9}$ &\\
                   	
	&	 R$_{in}$ (R$_{ISCO})$	&	$-$	&	 $-$ 	&	 $-$ 	&	 $4.3_{-2.4}^{+1.2}$ 	&	$1.9_{-0.4}^{+0.3}$ & \\
 
 &	 $\rm A_{Fe}$	&	$-$ &	 $-$ 	&	 $-$ 	&	 $1.8_{-0.4}^{+0.7}$ 	&	$4.95 \pm 0.0002$ &	 \\
  &	 q	&	$-$ &	 $-$ 	&	 $-$ 	&	 $3^f$ 	&	 $3^f$ &	 \\
   &	 $a^\ast$	&	$-$ &	 $-$ 	&	 $-$ 	&	 $0.259^{f}$ 	& $0.259^{f}$	&	 \\
 &	 $log{\xi}$ (erg cm s$^{-1}$) 	&	$-$	&	 $-$ 	&	 $-$ 	&	 $2.6_{-0.12}^{+0.04}$ 	&	$3.2 _{-0.1}^{+0.2} $ & \\

 &	 $log{N}$ (cm$^{-3}$) 	&	$-$	&	 $-$ 	&	 $-$ 	&	 $18.6_{-0.3}^{+0.4}$ 	&	$19.5_{-0.3}^{+0.2}$ & \\
 
 &	 $Refl_{frac} $	&	$-$	&	 $-$ 	&	 $-$ 	&	 $-1^{f}$ 	&	$-1^{f}$ & \\

 &	 norm ($\times10^{-3}) $	&	$-$	&	 $-$ 	&	 $-$ 	&$1.0_{-0.1}^{+0.2}$ 	&	$2.8\pm0.4$ &	& \\
\hline 
&	 Flux$_{1-79~keV}^a$ 	&	 &	 &	 	&	 $5.15 \pm 0.01 $ 	&	$10.20 \pm 0.01$ & \\
                   	
&	 Luminosity$_{1-79~keV}^b$ 	&	  	&	  	&	  	&	 $2.210 \pm 0.005$ 	& $4.390 \pm 0.005$ &	\\
\hline                   																
&	 $\chi^2$/dof  &	4320/2921	&	3455/2920             	&	2984/2917	&	2930/2914         &  1731/1599     & 	\\
\hline					
		\multicolumn{10}{l}{$^a$ : Unabsorbed flux in the units of $10^{-9}$ \erg.}\\
    \multicolumn{10}{l}{$^b$ : X-ray luminosity in the units of $10^{37}$ \lum, assuming a source distance of 6 kpc.}\\
      \multicolumn{10}{l}{$^f$ : Frozen parameters.}\\
	\end{tabular}}
\end{table*} 
\begin{figure*}
\centering
\includegraphics[width=0.55\columnwidth,angle=270]{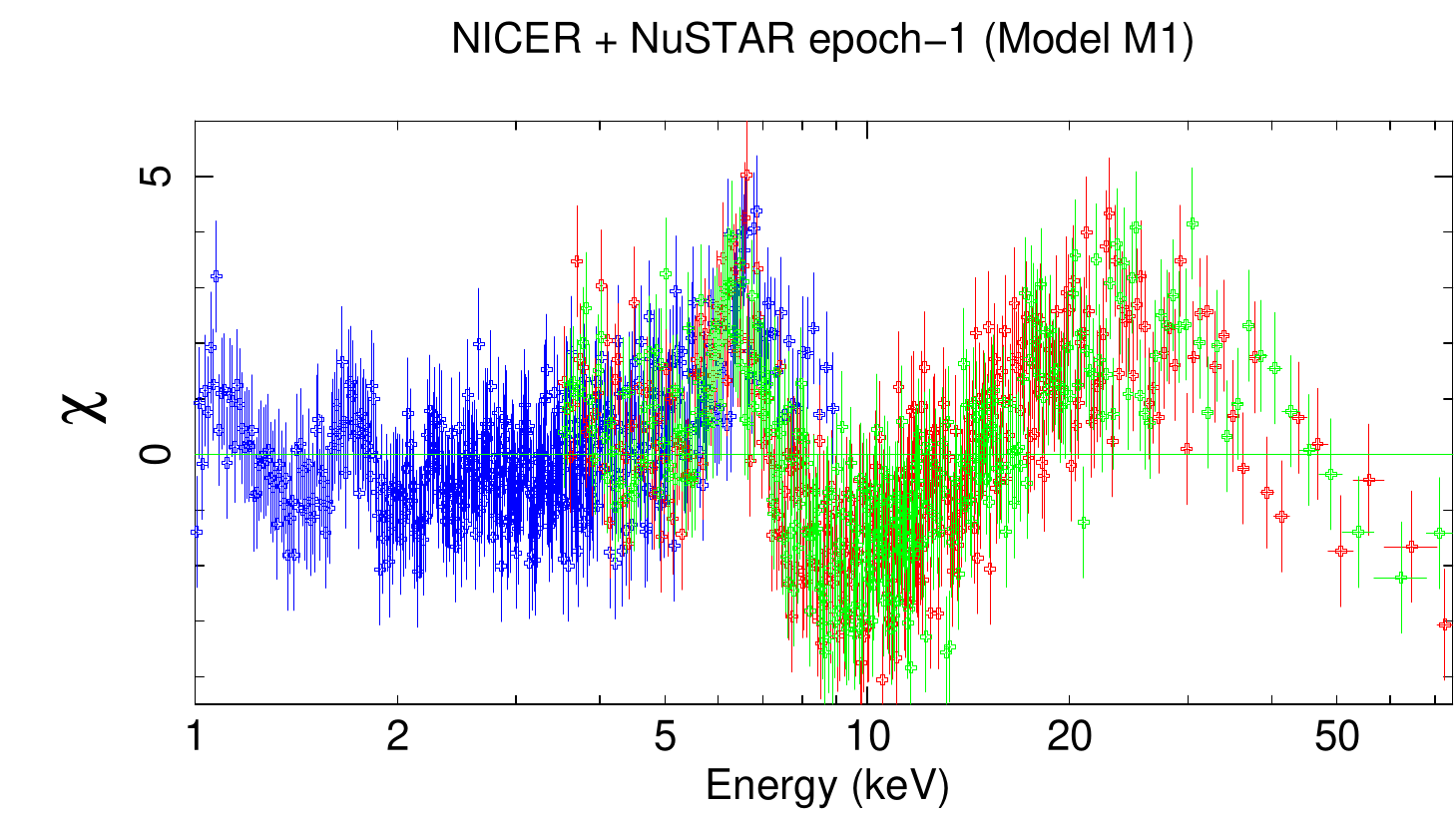}
\includegraphics[width=0.55\columnwidth,angle=270]{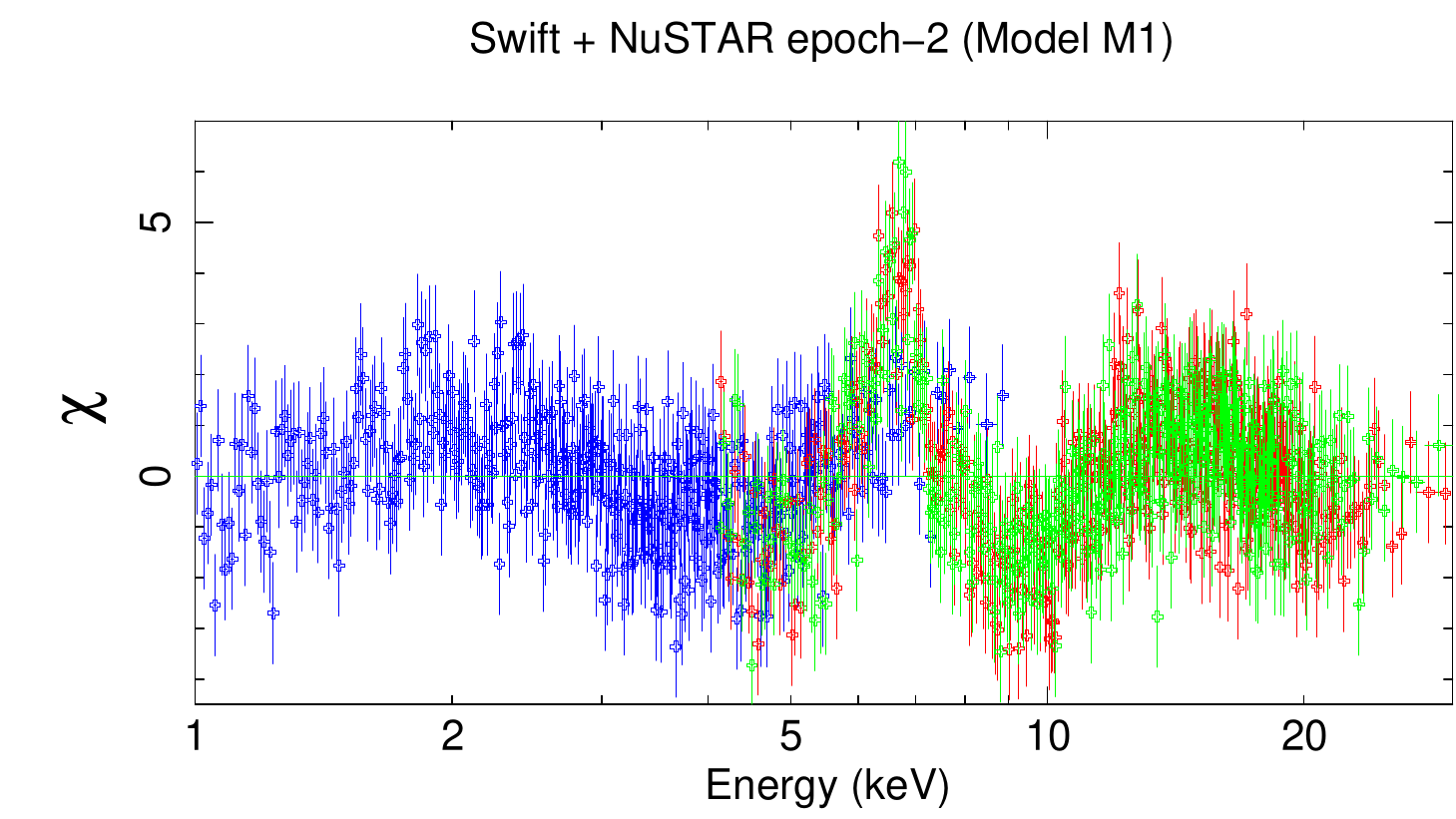}
 \caption{The residual plots of the broadband spectra of \nicer{}, \swift{} and \nustar{} of \src{} using the model \texttt{tbabs$\times$nthcomp}. A broad iron line at 6.4 keV and a Compton hump above 20 keV is evident.} 
\label{fig:delcchi}
\end{figure*}

\begin{figure*}
\centering
 \includegraphics[width=0.75\columnwidth, angle=270]{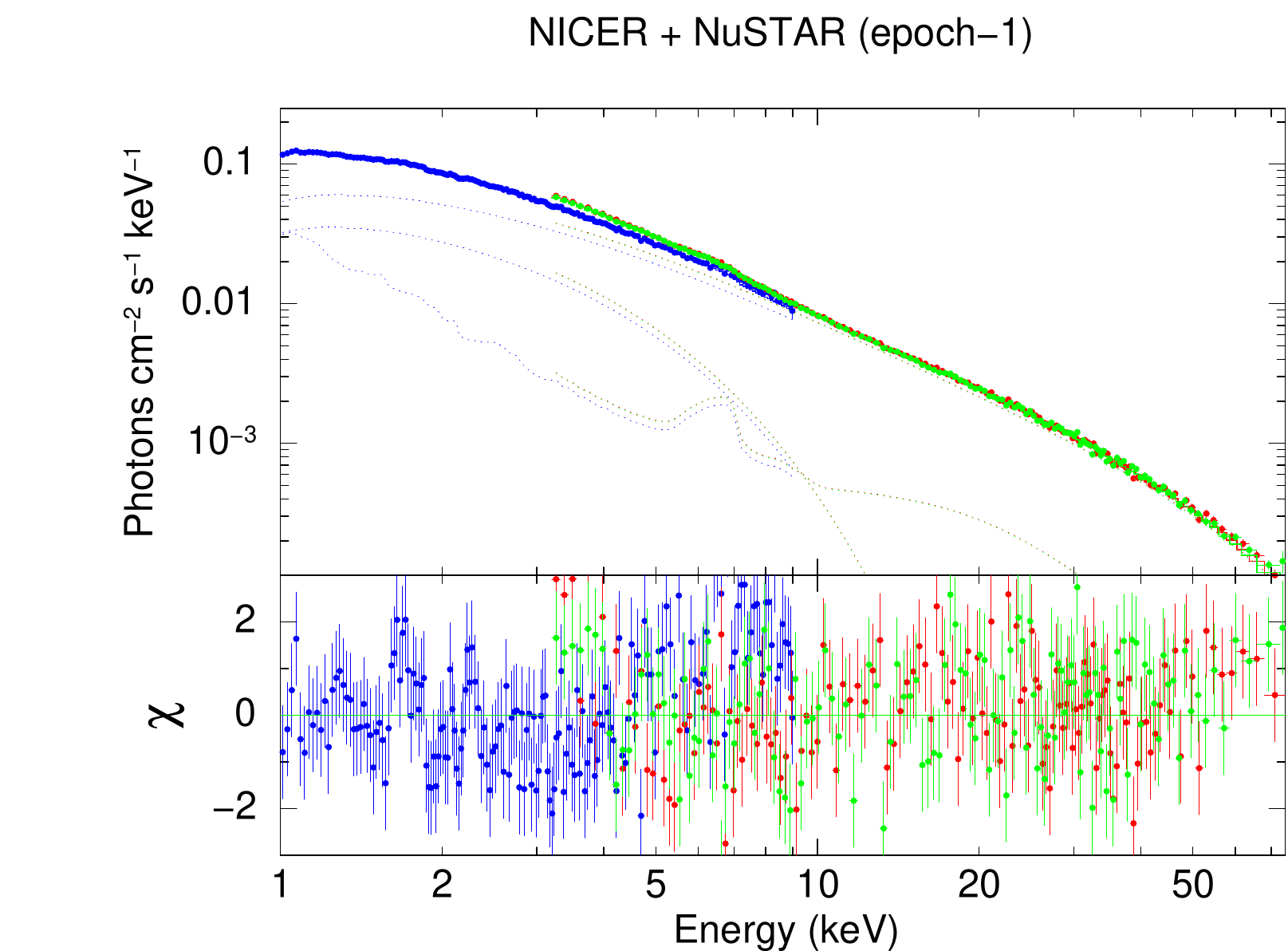}
 \includegraphics[width=0.75\columnwidth, angle=270]{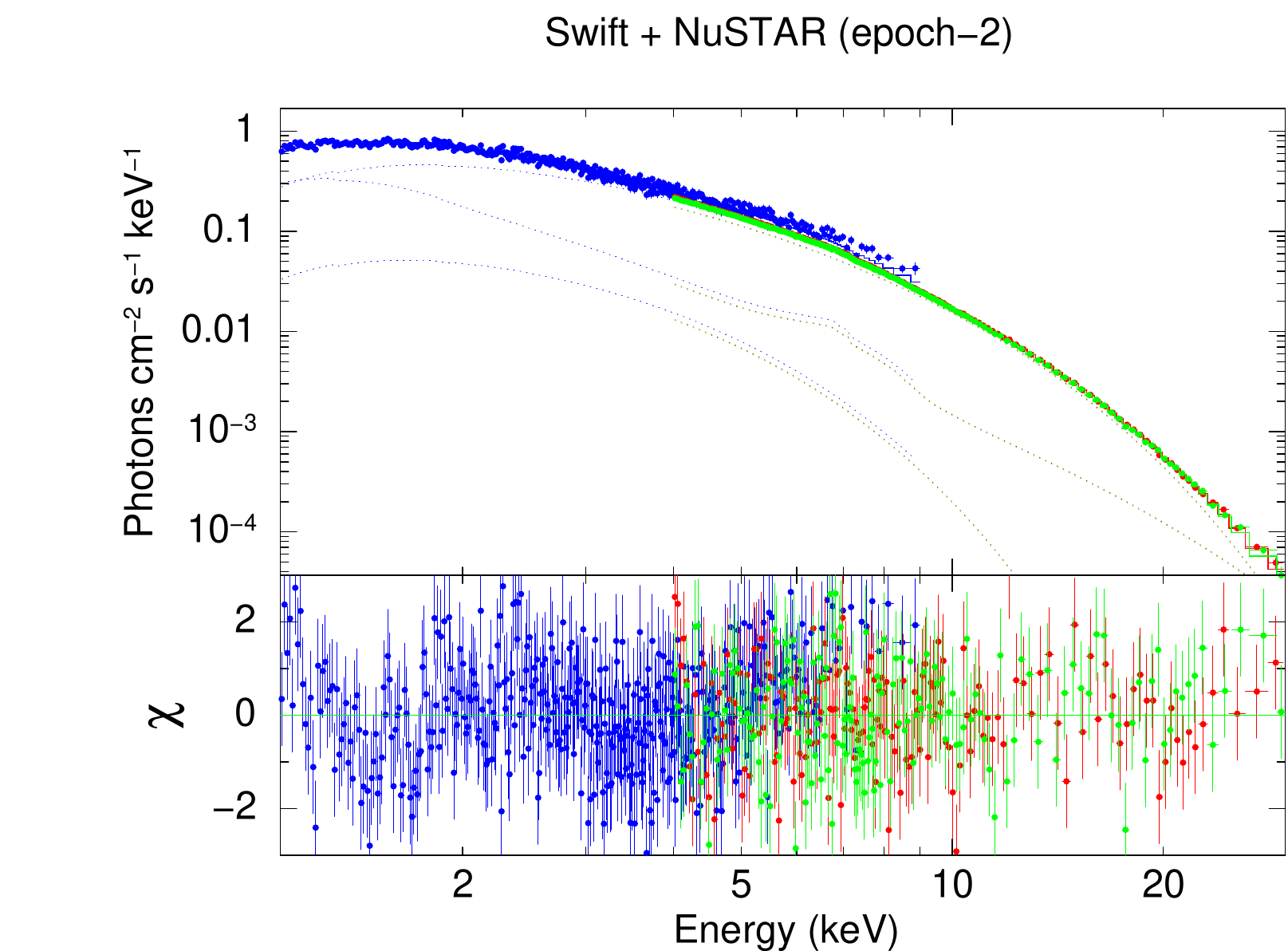}
 \caption{Best-fit broadband \nicer{} and \nustar{} (obs-1) spectra (left panel) of \src{} fitted with model M4 -- \texttt{tbabs$\times$(nthcomp+diskbb+relxillCp)}. The right-hand side panel shows the broadband best-fit spectra for the \swift{}/XRT and \nustar{} (obs-2) spectra with the model M4. The bottom panel shows corresponding residuals. Data is rebinned to present in the plot.}
\label{fig:spec}
\end{figure*}

\subsubsection{Modeling of the reflection features}

We approach modeling the broadband energy spectra systematically. Before moving on to the complex models, we first applied simple single continuum models including the single temperature blackbody component ({\tt bbodyrad} in {\tt XSPEC}), the multi-temperature disk blackbody component ({\tt diskbb} in {\tt XSPEC} notations: \citet{Mi84, Ma86}), thermal Comptonization component ({\tt nthcomp} in {\tt XSPEC} notations: \citep{Zd96, Zy99} and the power law model, modified by the {\tt TBabs} model. Modeling with these single continuum models resulted in reduced chi-square ($\chi^2$) values $>$ 1.5, indicative of a poor data description. In the next step, we introduced moderately complex three-component models such as {\tt TBabs×(diskbb + nthComp) or TBabs×(diskbb + cutoffpl)} to fit the broadband spectra. We observed a slight improvement in the reduced $\chi^2$. In the residuals (see Fig.~\ref{fig:delcchi}), we detected X-ray reflection features, including emission features in the 5--8 keV range, a photoelectric absorption dip around 10 keV, and a back-scattering Compton hump above 20 keV \citep{Fa89, Fa00, Mi07}. The reflection feature is evident irrespective of the choice of continuum components. Physically, these reflection features are most likely due to relativistic reflection from the accretion disk. The {\tt relxill/rexillCp} is a self-consistent broadband reflection model from the {\tt relxill} (v2.3: \citet{Ga14, Da14}) model family, which describes relativistic reflection due to an incident cut-off power-law or {\tt nthcomp} continuum. The broad-band spectrum shows a broad iron line around 6.4 keV followed by a Compton hump, which indicates the reflection feature. We further used a relativistic reflection model to probe this feature. The primary source is assumed to have an emissivity given by Index1 (q1) and Index2 (q2). The reflection model is further defined by parameters such as the photon index of the illuminating radiation, the ionization parameter ($\xi$) at the disk’s surface, and the abundance of Fe relative to its solar value. We fitted the broadband spectra with the model {\tt TBabs$\times$(nthcomp + diskbb + relxillCp)} and found a significant improvement in the fit quality (reduced $\chi^2 \sim$ 1). The best-fit broadband spectra are shown in Fig.~\ref{fig:spec} for two different \nustar{} observations. The bottom panel of Fig.~\ref{fig:spec} represents the residual, which indicates that the model combination can address all the features in the broadband spectra. The best-fit spectral parameters are summarized in Table~\ref{tab:fitstat2}.

The {\tt relxillCp} model includes the thermal Comptonization model {\tt nthcomp} as an illuminating continuum.  The emissivity index is fixed at q1 = q2 = 3, which eliminates the significance of $R_{br}$. During the modeling of the broadband spectra, we fix the outer radius of the accretion disk at 400 gravitational radii ($\rm R_g = GM/c^2$). The dimensionless spin parameter ($a^\ast$) is frozen at 0.259 \citep{Ki16,Lu17}. The reflection fraction of the {\tt relxillCp} model is fixed at a negative value (-1). Therefore, the {\tt nthcomp} describes the direct coronal emission, and the {\tt relxillCp} represents the reflected component. The power-law photon index ($\Gamma$) and the electron temperature ($kT_e$) of the {\tt relxillCp} model are tied with the {\tt nthcomp} model. Apart from the parameters mentioned above, using the {\tt relxillCp} model, we also determine various relativistic reflection and the continuum parameters, including the iron abundance ($A_{Fe}$), $\Gamma$, the accretion disk ionization parameter (log $\xi$), the disk inclination angle ($i$), and the inner-disk radius ($R_{in}$). The broadband spectral fitting result indicates that the disk is moderately ionized with the ionization parameter  log$\xi \sim$2.6 erg cm s$^{-1}$. The iron abundance ($\rm A_{Fe}$), inner disk radius R$_{in}$, and disk inclination $i$ are found to be $\sim$1.8, $\sim22 R_g$, and $33.1_{-3.2}^{+5.8}$ $^{\circ}$ , respectively.
The density of the accretion disk ($\log N$) is found to be $\sim18.6$ cm$^{-3}$ for the first \nustar{} observation (obs-1). 

For the second \nustar{} observation (obs-2), the broadband \nustar{} and \swift{}/XRT spectra are investigated for the reflection signatures. The broadband spectral fitting of \swift{} and \nustar{} spectra in the 1--30 keV range using the above model indicates a reflection feature including a broad iron line at $\sim$6.2 keV of width $\sim$1 keV and a Compton hump around 20 keV. The best-fit spectra and fitting parameters are shown in Fig.~\ref{fig:spec} and Table~\ref{tab:fitstat2}, respectively. The best-fit spectral fitting provides an inner disk radius of R$_{in}$ $\sim9.8 R_g$, and the disk inclination of $\sim20^{\circ}$. The density of the accretion disk and ionization parameter are found to be $\log N$ $\sim 19.5$ cm$^{-3}$ and $\log \xi \sim$3.2 erg cm s$^{-1}$, respectively. The iron abundance is found to be high ($A_{Fe}\sim 5$) for this observation. The reported errors in the spectral parameters are estimated using the Markov Chain Monte Carlo {\tt MCMC} approaches, and the {\tt MCMC} chain corner plot is shown in Fig.~\ref{fig:corner}. The {\tt MCMC} simulation is performed with the reflection model ({\tt relxillCp}) using the Goodman-Weare algorithm \citep{Good2010} for a chain length of 500000 and 10 walkers. We chose to discard the first 50000 steps, assuming them in the `burn-in' or `transient' phase.

\section{Discussion}
\label{dis}
We study the temporal and spectral properties of \src{} with \nicer{}, \swift{}, and \nustar{} during its 2024 outburst. During these observations, a total of six thermonuclear X-ray bursts were detected from \src{}. The burst profiles show strong energy dependence. Initially, we performed the spectral modeling of each pre-burst spectrum. The pre-burst spectra are modeled with a model consisting of an absorbed power law and a blackbody component. The best-fit pre-burst spectral parameters are used to account for the contribution of persistent emission during the modeling of the burst spectra. To investigate the evolution of spectral parameters during the thermonuclear bursts, we performed time-resolved spectroscopy for each burst detected in the \nicer{} and \nustar{} observations of \src{}. The burst time-resolved spectra are modeled using the variable persistent emission method ($f_a$ method), which better describes the spectral excess observed after the main blackbody component. The scaling parameter $f_a$ reached a maximum value of $\sim$3 to 5 during the peak of these bursts, as observed with \nicer{} and \nustar{}. 

Using a large sample of bursts observed with \nicer{} between 2017 July and 2021 April, the time-resolved burst spectral study of \src{} reported a peak value of $f_a$ $<5$ for non-PRE bursts \citep{Gu22}. However, in the case of the PRE bursts, the $f_a$ parameter increased to a maximum value of $\sim$10 (for details, see Fig.~4 \citep{Gu22}). There are several possible explanations for the deviation of the burst emission from a pure blackbody emission. The deviation is likely due to the atmospheric effect \citep{Ozel2013}. The atmosphere model of NS predicts a broad X-ray spectrum rather than a pure blackbody \citep{Su11, Ma14}. The relativistic effect may also have an impact on broadening the features in the X-ray spectrum of pulsars \citep{Bau2015}. The $f_a$ method does not provide a detailed physical explanation of the increment of the pre-burst emission. However, the enhancement can be associated with increased mass accretion rate to the NS due to the Poynting–Robertson drag \citep{Walker1992, Worpel2013}. The reflection of burst photons from the accretion disk is also anticipated (see, e.g., \citet{Gu22}). Considering this, we attempted to investigate the contribution of disk reflection during the bursts. The soft excess emission during the burst can be modeled using a disk reflection model {\tt relxillNS}. The reflection modeling approach provides a self-consistent and physically motivated explanation. This approach was adopted to model the time-resolved burst spectra for 4U~1820--30 \citep{Ke18, J24}, 4U~1636--536 \citep{Zh22}, 4U~1730--22 \citep{Lu23}, and SAX~1808.4--3658 \citep{Bu21}. We found that the reflection component can contribute $\sim$20\% of the total burst emission during \nicer{} burst. The blackbody temperature and blackbody flux obtained from the $f_a$ model are higher than those of the reflection model. During the burst, the flux of the reflection component is found to be strongly correlated with the blackbody flux as observed with \nicer{}. Due to poor statistics in \nustar{} data, the reflection modeling does not result in any significant improvement over the $f_a$ modeling approach.  

The time-resolved spectroscopy does not show any evidence of photosphere radius expansion during any of the bursts detected during the 2024 observations. The blackbody temperature reached a peak value of $\sim$3 keV and the bolometric flux was $\sim 5\times10^{-8}$ erg cm$^{-2}$ s$^{-1}$ at the peak of the \nicer{} burst. This peak flux is almost half of the Eddington limited flux of $F_{Edd} = 10.44 \times 10^{-8}$ erg s$^{-1}$ cm$^{-2}$ from the PRE-bursts \citep{Gu22}.  

Based on the local accretion rate, different thermonuclear ignition regimes can be probed. Assuming spherical accretion, the luminosity can be written as \citep{Ga06, Joh20}, 
\begin{equation}
L = 4\pi d^2 F \zeta = \frac{z \dot{M} c^2 }{(1+z)^3}  
\end{equation} 
where $\dot{M}$ is the mass accretion rate as measured at the NS surface, $\zeta$ is the anisotropy factor, and $F$ is the pre-burst flux. The mass accretion rate is estimated using the average \nustar{} pre-burst flux (1--79 keV range) of $\sim5.75\times$10$^{-9}$ erg cm$^{2}$ s$^{-1}$ and assuming the source distance of $\sim$6 kpc.

\begin{equation}
\dot{M} = \frac{(1+z)^3 4\pi d^2 F}{z \times c^2} \zeta ~~~~ \rm gm ~s^{-1}
\end{equation}

\begin{equation}
\dot{M} = 4.3 \times 10^{-10} \times \frac{(1+z)^3 \times \zeta}{z} ~~~~ M_\odot~~yr^{-1}
\end{equation}
Assuming an isotropic condition $\zeta=1$, for a typical NS of mass 1.4 M$_\odot$ and radius 10 km, the gravitational redshift on the surface of the NS is $(1+z) =\left(1-\frac{2GM}{c^2R}\right)^{-\frac{1}{2}}\simeq1.3$ and the mass accretion rate ($\dot{M}$) is estimated to be nearly $3.2\times$10$^{-9}$ $M_\odot$ yr$^{-1}$. For a typical NS, the approximate value of the mass accretion rate for stable hydrogen burning is $\sim3 \times$ 10$^{-10}$ $M_\odot$ yr$^{-1}$ \citep{Fu81}. During the pre-burst duration of \src{}, the mass accretion rate was higher than the required rate for stable hydrogen burning as estimated by \citep{Fu81}.

The theoretical ignition models of hydrogen and helium-burning bursts can be utilized to probe the evolution of burst properties with accretion rates for different systems. For stable hot-CNO cycle hydrogen burning, the mass accretion rate is related to the Eddington mass accretion rate as $\dot{M}$ $\ge$ 0.01 $\dot{M}_{Edd}$ \citep{Cu04, Ga06, Ga08}. We estimated the mass accretion rate $\dot{M}$ $\simeq$ 0.18 $\dot{M}_{Edd}$ (assuming $\dot{M}_{Edd} = 1.8 \times 10^{-8} M_\odot yr^{-1}$), which is higher than the estimated rates by \citet{Cu04} for stable CNO cycle hydrogen burning. Similarly, the preburst mass accretion rate is estimated from \nicer{} data. The \nicer{} preburst luminosity is found to vary between (1.2--2.8) $\times~10^{37}$ erg s$^{-1}$ for 3 observations. The corresponding mass accretion rate is estimated to be (1.5--3.6) $\times~10^{-9}$ $M_\odot yr^{-1}$ ($\sim$ (0.09--0.2) $\dot{M}_{Edd}$). At this level of mass accretion rate, the hydrogen burns stably, and the helium burning drives the thermal instability at a density of $\ge$ 10$^{5}$ gm cm$^{-3}$. Based on the time to accumulate the critical column of fuel, the burst may be a ``pure He'' burst or ``mixed H/He'' burst \citet{Cu04}. However, the longer burst duration ($>$50 s) indicates that the burst may be powered by the H/He mixed fuel or H-rich fuel \citep{Ga21}.


\subsection{Reflection modeling of persistent emission}
The broadband continuum spectra can be modeled with an absorbed Comptonized component and multi-temperature blackbody model. While fitting the persistent spectra of Aql~X-1, we detected the presence of X-ray reflection features in the residuals. A broad iron line was observed in both \nustar{} observations, which is followed by a Compton hump at around 20 keV. To model the reflection feature, we used an additional reflection model component {\tt relxillCp}.  The best-fit results provide a disk inclination of $33.1_{-3.2}^{+5.8}$ $^{\circ}$ and $19.5_{-2.7}^{+3.9}$ $^{\circ}$ for \nustar{} obs-1 and obs-2, respectively, with corresponding inner disk radii of $\sim4.3$ and $\sim2$ times of R$_{ISCO}$. Our measurements of disk inclination agree (within errors) with the infrared photometry measurements ($i$ < 31$^{\circ}$: \citet{Garcia1999}). The accretion disk is found to be moderately ionized with $\log \xi \sim$(2.6--3.2) erg cm s$^{-1}$ with high disk density of $\log N \sim(18.6-19.5)$ cm$^{-3}$. The electron temperature of the Comptonization component for the first \nustar{} observation was around 17 keV, which is much lower than during the second \nustar{} observation. This may be related to the evolution of the spectral state of the source. This kind of evolution in electron temperature is also reported for different LMXBs (for example, SAX~J1808.4--3658, \citet{Salvo2019}). X-ray reflection has been observed in different AMXPs with modern/high-resolution telescopes, including {\it XMM-Newton}, {\it NuSTAR}, and {\it INTEGRAL} \citep{Pa10, Pa13a, Pa13b, Pa16}. The inner disk radius is found to be relatively small along with the moderately broad iron line \citep{Pa10, Pa13a, Pa13b, Pa16}. But exceptions exist; for example, in a study of SAX~J1748.9--2021, \citet{Pi16} found a relatively high luminosity, corresponding to 25\% of the Eddington luminosity (L$_{Edd}$). The reflection features are not always detected in AMXPs, e.g, IGR~J16597--3704 \citep{Sa18a}, IGR~J17379--3747 \citep{Sa18b}, XTE~J1807--294 \citep{Fa05} and XTE~J1751--305 \citep{Mi03}.

The optical depth of the corona can be estimated using the electron temperature ($kT_e$) and the photon index ($\Gamma$) using the following equation \citep{Zd96}, 

\begin{equation}
    \tau_{e} = \sqrt{\frac{9}{4} + \frac{m_{e}c^{2}}{kT_e} \frac {3}{(\Gamma + 2) (\Gamma - 1)}} - \frac {3}{2}
\end{equation}

Using the best-fit spectral parameters for the {\tt nthcomp} model, the optical depth of the corona is estimated to be  $\tau_{e}$  $\sim$4.2 during \nustar{} obs-1. Similarly, for \nustar{} obs-2, the optical depth is estimated to be $\sim$8.4. The seed photons are up-scattered by the hot electrons in the Compton cloud and gain energy \citep{Su80, Su85}. The relative energy gained by photons during inverse Compton scattering can be estimated by the Compton-y parameter \citep{Ry79}, 

\begin{equation}
    y = \frac {4kT_e\tau_e^{2}}{m_ec^2}
\end{equation}

 Based on the best-fit spectral parameters, the y-parameter is estimated to be nearly 2.4 and 1.4 for \nustar{} obs-1 and obs-2, respectively.

The self-consistent reflection model, {\tt relxillCp}, describes the observed reflection feature. This model provides the inner disk radius of $\sim4.3$ R$_{ISCO}$. For a rotating neutron star with dimensionless spin parameters $a^\ast$ = 0.259, $\rm R_{ISCO}$ is defined as \citet{Va04}, 

\begin{equation}
    R_{\rm ISCO} = \frac{6GM}{c^2} (1-0.54a^\ast) = 5.16 \frac{GM}{c^2} = 5.16~ R_g
\end{equation}
where, $G$ is the gravitational constant, $M$ is the mass of the neutron star and $c$ is the speed of light. From the best-fit spectral parameters, $\rm R_{in} = 4.3~R_{ISCO} = 22.2~ R_g = 46 ~km$ for \nustar{} obs-1. Similarly for the second \nustar{} observation, the inner disk radius is estimated to be $\rm R_{in} = 1.9~R_{ISCO} = 9.8~ R_g = 20.3 ~km$. The measurements of the inner disk radius indicate that the disk may be moderately truncated away from the NS surface. For NS LMXB, the accretion disk is supposed to be truncated at a moderate radius because of the pressure of the magnetic field of the neutron star \citep{Ca09}. Assuming the disk is truncated at the magnetosphere radius, the magnetic field strength can be estimated. Using the value of the inner disk radius \nustar{} obs-1, the magnetic dipole moment ($\mu$) and the magnetic field can be estimated using the following relation from \citet{Ib09}, 

\begin{math}
 \mu = 3.5 \times 10^{23} k_A^{-7/4} x^{7/4} \left(\frac{M}{1.4 M_\odot}\right)^2 \times     \left(\frac{f_{\rm ang}}{\eta} \frac{\rm F_{bol}}{10^{-9} \rm erg~cm^{-2}~s^{-1}}\right)^{1/2} \frac{D}{3.5 \rm kpc}
\end{math} \\
where, $\rm f_{ang}$ is anisotropy correction factor and $\eta$ is accretion efficiency, and $k_A$ is geometric coefficient. $M$ is the mass of the NS and $\text F_{bol}$ is the unabsorbed bolometric flux. We used 1--79 keV flux as bolometric flux, which is $\text F_{bol} \sim 5.2 \times 10^{-9}$ erg cm$^{-2}$ $s^{-1}$. The scaling factor $x$ can be estimated from $\text R_{\rm in} = \frac{xGM}{c^2}$. Assuming $f_{\rm ang} = 1$, $\eta = 0.1$, and $k_A = 1$ \citep{Ca09}, the magnetic dipole moment of the NS can be estimated to be $\mu = 9.6 \times 10^{26}$ G cm$^3$. The corresponding magnetic field strength at the poles of NS of radius 10 km is $B~ \sim 1.9 \times 10^9$ G. Similarly, for the \nustar{} obs-2, the magnetic dipole moment and the magnetic field are estimated to be nearly $3.3\times10^{26}$ G cm$^{3}$ and $0.6\times10^{9}$ G, respectively. The estimated magnetic field is consistent with the typical limit of other NS LMXBs \citep{Mu15, Lu16}.

\section{Summary and conclusions}
\label{con}
We studied the broadband spectral and timing properties of \src{} during its outburst in 2024 with \nicer{}, \swift{}, and \nustar{}. We detected 6 type-I X-ray bursts with \nicer{} and \nustar{}. The time-resolved spectra of bursts are modeled using an absorbed blackbody component, and the non-thermal persistent emission is modeled with the scaled background ($f_a$) approach. A soft excess was evident in the time-resolved burst spectroscopy, which indicates a deviation from the pure blackbody emission, possibly due to the enhanced mass accretion rate onto the NS. Alternatively, the soft excess can be modeled with the disk reflection model, and the contribution from the reflection model is found to be $\sim$20\% of the total emission as observed during the \nicer{} burst. During the burst, the flux of the reflection component is found to be strongly correlated with the blackbody component flux. The bursts did not show any evidence of PRE. The preburst mass accretion rate is estimated to vary between $\sim$ (0.09-0.2) $\dot{M}_{Edd}$ for all six bursts. We also detected X-ray reflection features, including a broad iron line at 6.4 keV and a Compton hump at around 20 keV with broadband \nustar{}, \nicer{}, and \swift{} spectral study. The evolution of spectral parameters and disk properties are also investigated using the X-ray reflection spectral study at two different outburst phases. The accretion disk is found to be moderately ionized ($\log \xi \sim(2.6-3.2)$ erg cm s$^{-1}$ ) with a high disk density of $\log N \sim(18.6-19.5)$ cm$^{-3}$.   Assuming that the inner disk is truncated at the magnetosphere boundary, the magnetic field strength at the poles of the neutron star is estimated to be $\sim(0.6-1.9) \times 10^9$ G.


\section*{Acknowledgements}
 We thank the referee for useful comments, which helped in improving the paper. The research work at the Physical Research Laboratory, Ahmedabad, is funded by the Department of Space, Government of India. This research has made use of data obtained with {\it NuSTAR}, a project led by Caltech, funded by NASA, and managed by NASA/JPL, and has utilized the {\tt NUSTARDAS} software package, jointly developed by the ASDC (Italy) and Caltech (USA). MM is thankful to Ajay Ratheesh (INAF/IAPS) for the discussion on reflection models and the comments on the draft. This research has made use of the {\it MAXI} data provided by RIKEN, JAXA, and the {\it MAXI} team. We acknowledge the use of public data from the {\it NuSTAR}, {\it Swift}, and {\it NICER} data archives. We thank the {\it NuSTAR} SOC Team for making this ToO observation possible. We are also thankful to the {\it NICER} team for the continuous monitoring of the source.

\section*{Data Availability}
The data used for this article are publicly available in the High Energy Astrophysics Science Archive Research Centre (HEASARC) at \\
\url{https://heasarc.gsfc.nasa.gov/db-perl/W3Browse/w3browse.pl}.

\appendix

\section{MCMC corner plot}
This figure shows the MCMC corner plot for model M4 at two different phases of the outburst.
\begin{figure*}
\centering
 \includegraphics[width=0.9\linewidth]{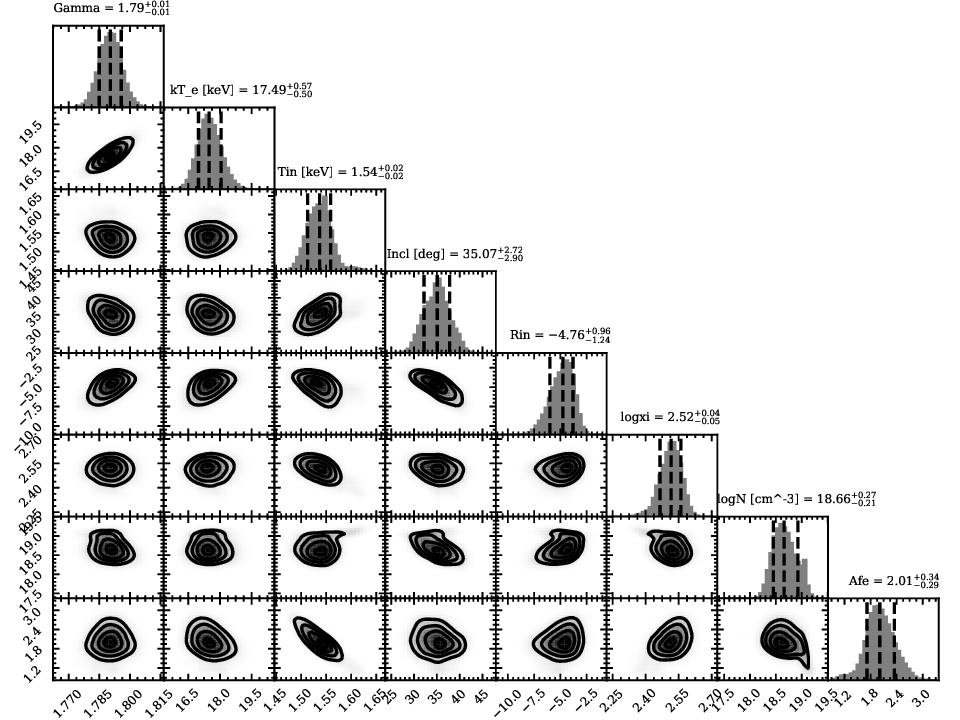}
  \includegraphics[width=0.9\linewidth]{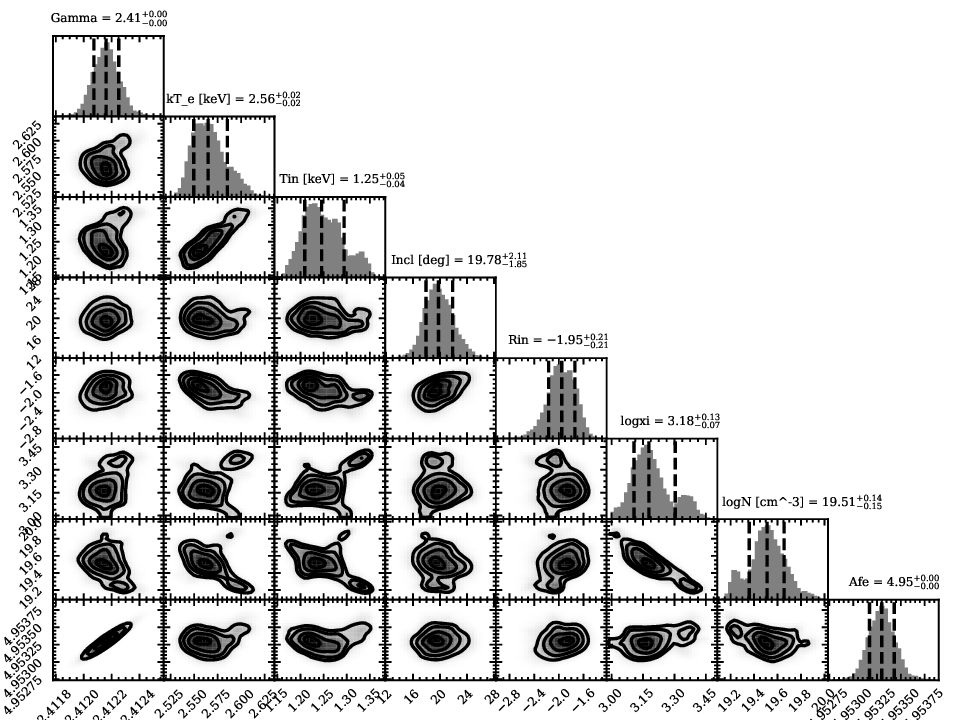}
 \caption{MCMC chain corner plot is shown for the joint \nustar{} obs--1 + \nicer{} (top panel) and \nustar{} obs--2 + \swift{} (bottom panel) best-fitting model M4 of \src{}.}
\label{fig:corner}
\end{figure*}




\end{document}